\documentclass{aastex}          
\usepackage{spr-astr-addons}    

\begin{document}
%
  \title{A model of neutron-star--white-dwarf collision for fast radio bursts}

\shortauthors{Liu}

\author{Xiang Liu\altaffilmark{1,2}}


\altaffiltext{1}{Xinjiang Astronomical Observatory, Chinese
Academy of Sciences, 150 Science 1-Street, Urumqi 830011, PR
China}

\altaffiltext{2}{Key Laboratory of Radio Astronomy, Chinese
Academy of Sciences, Nanjing 210008, PR China}



\begin{abstract}

Fast radio bursts (FRBs) with unknown origin emit a huge luminosity (about 1\,Jy at 1\,GHz) with a duration of milliseconds or less at extragalactic distances estimated from their large dispersion measure (DM). We propose herein a scenario for a collision between a neutron star (NS) and a white dwarf (WD) as the progenitor of the FRBs by considering the burst duration scaling to the collision time and the radio luminosity proportional to the kinetic energy of the collision. The relations among the observed flux density, pulse width, and the DM are derived from the model and compared with the statistical results from the observed FRBs. Although the sample is quite small, we tentatively report a nearly inverse-square correlation between the observed peak flux density and the DM excess, which is an consequence of the assumption that the DM excess (i.e. that not due to our Galaxy) is dominated by the intergalactic medium. We also tentatively note a correlation among the duration of the FRB and the DM excess (possibly interpreted as due to the broadening of the signal in the intergalactic medium) and a correlation among the duration of the FRB and the flux density (shorter burst should be brighter), both roughly in agreement with the proposed model.

\end{abstract}


\section{Introduction}

The first fast radio burst (FRB) was found by Lorimer et al. (2007). Today, 34 FRBs have been discovered (Keane et al. 2012; Thornton et al. 2013; Burke-Spolaor \& Bannister 2014; Spitler et al. 2014; Petroff et al. 2015; Ravi et al. 2015; Champion et al. 2016; Masui et al. 2015; Keane et al. 2016; Ravi et al. 2016; Petroff et al. 2017; Caleb et al. 2017; Bannister et al. 2017; Shannon et al. 2017; Bhandari et al. 2018; Price et al. 2018; Oslowski et al. 2018a,b; Farah et al. 2018) and catalogued\footnote{http://www.frbcat.org} (Petroff et al. 2016). The FRBs  show strong radio pulses on the level of 1\,Jy with millisecond  durations and with large dispersion measures (DMs) well exceeding the Galactic DM which is estimated by using the models of Cordes \& Lazio (2002) and Yao et al. (2017), suggesting an extragalactic origin of the FRBs. Only  FRB121102 has been found to show repeating radio pulses (Spitler et al. 2016; Scholz et al. 2016; Law et al. 2017), and it has been localized to a low-metallicity star-forming dwarf galaxy at z=0.19 (Chatterjee et al. 2017; Tendulkar et al. 2017). Others have not been found to repeat, and no emission from other wavelengths is found (Scholz et al. 2017).

 The short millisecond duration implies an emission-region size not larger than the magnetosphere of a neutron star (NS). The luminosity of the FRBs can be as high as \({\sim}1.2\times10^{42} \) erg/s for a burst flux density of 1 Jy at 1 GHz originating at 1 Gpc. The brightness temperature of FRBs, e.g., for the FRB121102 at 1 Gpc, is estimated to be \({\sim}10^{35}\) K (Lyutikov 2017), which is many orders of magnitude greater than the inverse-Compton limit (\({\sim}10^{12}\) K), so the radio radiation must be from a coherent mechanism. The origin of the FRBs is currently unknown, with non-cataclysmic models suggesting giant flares from magnetars (see, e.g., Popov \& Postnov 2010; Thornton et al. 2013; Pen \& Connor 2015) and supergiant pulses from extragalactic neutron stars (see, e.g., Cordes \& Wasserman 2016) as possible progenitors. The cataclysmic progenitor theories include NS mergers (Totani 2013) and ``blitzars'' which occur when a neutron star collapses to a black hole (BH) (Zhang 2014; Falcke \& Rezzolla 2014), and the collapse of a strange star crust (Zhang et al. 2018). A superluminous supernova interacting with its environment is also proposed as the origin of FRBs (Piro 2016). See Katz (2016, 2018) for a brief review. However, a NS-NS merger might not account for the FRBs, because the expected radio emission would be too weak, as seen in the first NS-NS collision event found in GW170817/GRB170817A (Abbott et al. 2017a,b,c). Furthermore, the interaction involves a body (e.g., planet, asteroid) orbiting an extragalactic pulsar is also proposed for the origin of FRBs (Mottez \& Zarka 2014; Geng \& Huang 2015; Dai et al. 2016).

Recent studies suggest that the coherent synchrotron mechanism is disfavored for the FRBs because it would produce the required luminosity at excessively large frequencies (Ghisellini \& Locatelli 2018). The coherent curvature radiation from bunches of particles emitting coherently is capable of accounting for the FRBs luminosity in the GHz range (Kumar et al. 2017; Ghisellini \& Locatelli 2018), in which a magnetic field of \(10^{14} \) G and an electron density of \(n_{e}\sim10^{16}\text{--}10^{17}\text{ cm}^{-3}\) (Kumar et al. 2017) or \(B\ge10^{13} \)~G and \(n_{e}\sim10^{13}\text{--}10^{17}\text{ cm}^{-3}\) (Ghisellini \& Locatelli 2018) in the NS are required, which imply that the emitting particles are in an inner-magnetosphere not far from the surface of a NS. Kumar et al. (2017) also considered the magnetic reconnection (MR) in the NS as the energy source for FRBs. The MR might be due to the movement of the NS crust where the B fields are anchored. The distortion of the B-field lines in the crust or magnetosphere builds up over time until it reaches a critical state and becomes unstable and field lines are reconfigured (Kumar et al. 2017), as when energy is released by solar flares. This model should be applicable to the FRBs that have multiple outbursts such as FRB121102, but not to non-repeating FRBs (Lu \& Kumar 2018).

Thus, multiple giant pulses from young NSs, magnetars, and from the magnetic reconnections in NSs should be expected, which could explain the repeating FRB121102 but not the non-repeating FRBs. However, monitoring the non-repeating FRBs would still be valuable because it would allow us to search for if any repeating pulses on longer timescales.

 A genuine non-repeating FRB is more likely originated from cataclysmic events (e.g., the collision of two compact objects). In this paper, we propose that the collision between a neutron star and a white dwarf (WD) could produce an FRB, for the short collision time and the enormous quantity of electrons ejected from the WD into the magnetosphere of the NS during the collision. We thus investigate herein this scenario and compare the results that it implies with the statistical results of FRBs.

\section{Neutron-star--white-dwarf collision scenario for fast radio bursts}
\subsection{Description of scenario }

We explore a scenario of collision between a NS and a WD for FRBs in which we assume as an approximation that a tidal disruption does not significantly affect our model. In the simulations by Paschalidis et al. (2009, 2011), there are two fates in close NS-WD binaries, one is stationary mass transfer (SMT) from the WD across the inner Lagrange point onto the NS, that will evolve on a secular timescale for inspiral; the other is tidal disruption via unstable mass transfer (UMT), which will lead to a fast collision of the NS onto the WD in a hydrodynamical (orbit) timescale (Paschalidis et al. 2009, 2011). In the second fate, which is also called the ``head-on'' collision by Paschalidis et al. (2011), the NS may plunge into the WD and spiral into its center, forming a quasi-equilibrium configuration that resembles a Thorne-Zytkow
object (Thorne \& Zytkow 1977).

The fraction is approximately comparable for the NS-WD binaries which will undergo either UMT or SMT (Paschalidis et al. 2009), and the probability of the ``head-on'' collision is greater for the NS-WD binaries with higher WD/NS mass ratios (Paschalidis et al. 2011). Thus, the generic NS-WD merger rate may be dominated by the probability of the ``head-on'' collision of the NS-WD binaries through the UMT, because the SMT will take much longer time for the binaries to merge than the UMT (Paschalidis et al. 2009, 2011).

Millions of NS-WD binaries would be expected in our Galaxy and more than 146 NS-WD systems have been found (from pulsar catalog \footnote{http://www.atnf.csiro.au/research/pulsar/psrcat/}, Manchester et al. 2005). The event rate of FRBs is recently revisited with a larger sample of FRBs detected at Parkes to be an all-sky FRB rate of \(1.7^{+1.5}_{-0.9}\times 10^{3}/(4\pi sr)/\text{day}\) above a fluence of \(2\, \text{Jy ms}\) (Bhandari et al. 2018), and they computed the volumetric rate of FRBs for the FRB sample using the fluence complete rate as the basis and obtained the volumetric rates of FRBs of \(2000\text{--}7000\, \text{Gpc}^{-3}\text{yr}^{-1}\) out to a redshift of \(z\sim1\). This range of volumetric rates of FRBs are roughly consistent with the NS-WD merger rate of \(\sim0.5\text{--}1\times10^{4}\,\text{Gpc}^{-3} \text{yr}^{-1}\) (Thompson et al. 2009; Paschalidis et al. 2011). The NS-WD merger rate is also estimated to be \(\sim 10^{-3}\text{--}10^{-2}\) times the core collapse supernovae (CCSNe) rate (O'Shaughnessy \& Kim 2010; Margalit \& Metzger 2017), and the CCSNe rate is \(\sim7\times10^{5}\,\text{Gpc}^{-3}\text{yr}^{-1}\) in redshift \(z\sim1\) (Horiuchi et al. 2011). From these estimates, the NS-WD merger rate is \(700\text{--}7000\, \text{Gpc}^{-3}\text{yr}^{-1}\), which is also comparable to the FRB rate estimated by Bhandari et al. (2018) out to redshift of 1. The NS-WD merger rate may be higher than the NS-NS, NS-BH, and BH-BH merger rates (Mapelli \& Giacobbo 2018), because there are usually more low-mass stars in galaxies.

 In the close NS-WD systems, as mentioned above, at the final stage through the unstable tidal disruption the NS will quickly merges into the WD.  The size of a NS (typically 20~km size with 1.5 solar masses) is much smaller than that of a WD (typically the Earth size \(\sim\)10\,000~km with 0.6 solar mass) and the density of the NS is \({\sim}10^{8}\) times that of the WD. When the NS impacts on the WD, an electromagnetic burst from the magnetosphere of the NS is expected from the coherent curvature radiation (Cordes \& Wasserman 2016; Kumar et al. 2017; Ghisellini \& Locatelli 2018) due to the large amount of relativistic electrons ejected from the WD into the magnetosphere of the NS during the collision. The burst energy should mainly come from the kinetic energy of the NS, which is converted to the relativistic electrons ejected from the WD to the magnetosphere of the NS. The burst timescale is assumed to be scaled to that of the time interval from the start of impact until the NS fully merges with the WD, which equals the NS diameter \(D_{NS}\) divided by the average velocity \(\bar{V}\) of the NS relative to the WD during the collision. Detailed modeling and analysis based on the observed properties of the FRBs are presented in next subsection.

 The density of charge particles in the magnetosphere of a NS is \(0.07\times B_{z}/P \text{ cm}^{-3}\) (Goldreich \& Julian, 1969) with the axial component of the magnetic field \(B_{z}\) in Gauss and the pulsar period \(P\) in seconds. Thus, the density of charged particles decreases with distance from the NS surface.

The curvature radiation can be produced when charged particles move along curved magnetic field lines, as in the magnetosphere of a NS. Kumar et al. (2017) demonstrated that the isotropic equivalent luminosity of \(10^{43} \)~erg/s for FRBs in the observer frame, can be achieved by the sum of the coherent patches of curvature radiation in the magnetosphere of the NS, which contains an electron density of \({\sim}10^{16-17}\text{ cm}^{-3}\) greater than the Goldreich-Julian density (Goldreich \& Julian, 1969). Locatelli \& Ghisellini (2017) and Ghisellini \& Locatelli (2018) further investigated the absorption of the coherent curvature radiation (CCR) in various parameter spaces (e.g., for a curvature radius \(\rho=10^{4-7}\text{ cm}\), electron density \(n_{e}=10^{13-17}\text{ cm}^{-3}\), Lorenz factor \(\gamma=10^{1-3}\)) for both the monoenergetic particle distribution and the power-law particle distribution, and they suggested a strong radiation power peaked in the GHz range or higher and strong absorption below 100 MHz. In addition, they predicted an absent or very weak inverse-Compton emission at high energies. This model can be applied to the magnetosphere of a NS in collision with a WD, because the high electron density required for the CCR of FRBs can be efficiently added by the electrons ejected from the WD, and the B-field of the NS is probably enhanced as well in the collision with the WD. In addition, the burst energy can be obtained from the kinetic collision energy.

As the electron cooling time is of \({\sim}10^{-14}\text{s}\) in the CCR mechanism (Kumar et al. 2017; Ghisellini \& Locatelli 2018), the FRBs therefore require a continuous injection of energetic electrons in the milliseconds duration of FRBs. In the NS-WD collision model, the continuous injection of electrons from the WD into the magnetosphere of NS is possible during collision.

In the mean time, an electron-rich plasma may also be ejected and inflated into the outer-magnetosphere of the NS, with a lower electron density than that ejected into the inner magnetosphere of the NS. These plasma surrounding the NS may contribute to the DM of the FRBs. According to simulations made by Locatelli \& Ghisellini (2017) and Ghisellini \& Locatelli (2018) the opacity of the plasma or material (e.g., electrons, protons) ejected from the WD during a collision can be optically thin for the coherent curvature radiation at GHz frequencies to exit from the plasma with a density as high as \(n_{e}\sim10^{15}\text{ cm}^{-3}\).

The NS-WD collision model may be applicable to the non-repeating FRBs with the burst duration scaled to the merging duration, and the energy tipped from the kinetic energy of the collision, and the CCR mechanism applied. In contrast, the repeated FRB121102 may be explained by the NS-WD binary model proposed by Gu et al. (2016) in which a NS-WD system is not yet merged but transfers mass from the WD to the NS intermittently through the Roche lobe, leading to accreted magnetized plasma and triggering magnetic reconnection near the NS surface. There could be another option (besides the Gu et al.) to invoke NS+WD systems as progenitors of repeating FRB models, i.e. after the NS and WD merge, a new NS (probably a Thorne-Zytkow-like object, see Paschalidis et al. 2011, which might evolve into a magnetar) would form. Such a remnant NS would have a higher B field and an unstable surface, which would induce frequent magnetic reconnections, leading to a repeating FRB such as FRB121102. Additional work will be needed to explore this option, which is not in the focus of this paper.

\subsection[]{Model with statistical properties of fast radio bursts}

There are 34 FRBs up to May 28, 2018 in the catalog maintained by Petroff et al. (2016) and they appear to be distributed over the sky but not concentrated in the Galactic plane. However, note that the distribution of the FRBs also reflects the amount of time spent by the various surveys in given directions of the sky, as discussed by Keane (2016), Keane \& Petroff (2015) and references therein. The statistics of the 34 FRBs indicate a pulse width ranging from 0.35 to 26 ms, with a median of 2.94 ms and a mean of 4.95 ms, a peak flux density ranging from 0.2 to 128 Jy, with a median of 0.72 Jy and a mean of 8.78 Jy, a DM ranging from 176.4 to 2596.1 \(\text{pc/cm}^{3}\), with a median of 784.5 \(\text{pc/cm}^{3}\) and a mean of 897.5 \(\text{pc/cm}^{3}\). The DM excess (\(DM-DM_{\text{Galaxy}}\)) ranges from 139.4 to 2583.1 \(\text{pc/cm}^{3}\), with a median of 679.2 \(\text{pc/cm}^{3}\) and a mean of 795.6 \(\text{pc/cm}^{3}\). The NS-WD collision model attempts to explain the statistical properties of the non-repeating FRBs as described in the following.

The NS-WD collision model assumes the FRB duration to be scaled to the collision time (i.e., the time interval from the start of impact until the NS merges fully with the WD). Because the NS should decelerate during the impact, as a proxy for deriving the correct timescale of the event, we adopt the average velocity \(\bar{V}\) of the NS relative to the WD in the collision. For the NS diameter \(D_{\text{NS}}\), from the emitted pulse width \(W_{e}\) in the source frame and the observed pulse width \(W^{\text{iso}}_{o}\) of an FRB, we have \(W^{\text{iso}}_{o}=\kappa W_{e}(1+z)=\kappa\xi(D_{\text{NS}}/\bar{V})(1+z)\), where \(\kappa\) is the temporal broadening factor due to scattering by the ionized interstellar medium (ISM) and intergalactic medium (IGM), and \(\xi\lesssim 1\) is the scaling factor as \(W_{e}=\xi D_{\text{NS}}/\bar{V}\), which assumes an FRB emission region is comparable to or less than a NS size. We assume that the burst energy (\(L_{e}\)) is proportional to the mechanical kinetic energy of the NS when it impacts the WD, i.e. \(L_{e}=\varepsilon(0.5M_{\text{NS}}\bar{V}^{2})\), where \(\bar{V}=\xi D_{\text{NS}}/W_{e}\), \(\varepsilon<1\) is an efficiency, and we know \(L_{e}=4\pi D_{L}^{2}S^{\text{iso}}_{o}\nu_{e}W_{e}=4\pi D_{L}^{2}S^{\text{iso}}_{o}\nu^{\text{iso}}_{o}W^{\text{iso}}_{o}/\kappa\), \(D_{L}\) is the luminosity distance, \(S^{\text{iso}}_{o}\) the ``observed'' flux density from an isotropic source emission, \(\nu^{\text{iso}}_{o}\) is the observing frequency and \(\nu^{\text{iso}}_{o}=\nu_{e}/(1+z)\), \(z\) is the redshift. This gives:

\begin{equation}
S^{\text{iso}}_{o}=\frac{\varepsilon \xi^{2}\kappa^{3}M_{\text{NS}}D_{\text{NS}}^{2}(1+z)^{2}}{8\pi\nu_{o}^{\text{iso}}D_{L}^{2}}(W_{o}^{\text{iso}})^{-3}
\end{equation}

In this equation, the mass \(M_{\text{NS}}\) and diameter \(D_{\text{NS}}\) can be similar for different neutron stars, the scaling factor \(\xi\lesssim 1\), and \(\varepsilon\) is the energy conversion efficiency. The luminosity distance \(D_{L}\) can be quite different for individual FRBs at cosmological distances, which is reflected in the different \(DM\) excesses (as a proxy distance) observed for the FRBs. The temporal broadening factor \(\kappa\) may be quite different for the FRBs, if the IGM can be well approximated
by density inhomogeneities. We are interested in the observed quantities, for example, \(S_{o}\), \(W_{o}\), \(DM_{\text{excess}}\), and the relations among them, such as the inverse-power-law relation between the flux density and the width of FRBs in Eq. (1).

\begin{figure}
    \includegraphics[width=8.4cm]{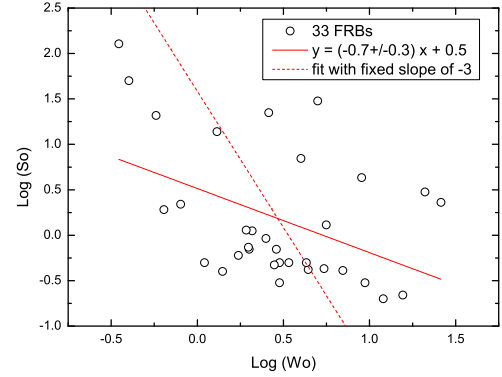}
    \caption{Log-log plot of observed peak flux density (\(S_{o}\) [Jy]) vs. observed pulse width (\(W_{o}\) [ms]) for 33 FRBs excluding the repeating FRB121102. The solid line is the best linear fit, and the dash line is the fit with the fixed index of \(-3\) implied by Eqs. (1) and (3).}
     \label{fig1}
  \end{figure}

To compare the relationship with the statistical result of the FRBs catalogued by Petroff et al. (2016), the flux density and width of 33 FRBs are plotted in Fig.~\ref{fig1}, where the repeated FRB 121102 is excluded because it may have different origins. In the log-log plot, the best linear fit gives a slope of \(-0.7\pm0.3\), suggesting a negative power-law correlation between observed flux density and width of the FRBs. However, that relation is not as steep as Eq. (1) would predict (i.e. a slope of \(-3\)). This difference could be due to the scattered data caused by several factors, e.g., (i) the luminosity distance \(D_{L}\) can differ significantly for the 33 FRBs; (ii) the temporal broadening \(\kappa\) may be quite different for the FRBs; (iii) the source might not emit isotropically, it could emit more in the magnetic polar region of the NS and be Doppler boosted as well, and the 33 FRBs may have quite different view angles with respect to their magnetic axis; and (iv) the sample may not be large enough to yield proper statistics.

By using Eq. (1), we could in principle estimate the distance \(D_{L}\) for the FRBs, if we assume a NS mass of 1.5 \(M_{\odot}\) and the NS diameter of 20 km (\"Ozel \& Freire 2016). With the observed \(S^{\text{iso}}_{o}\) and  \(W^{\text{iso}}_{o}\) at \(\nu^{\text{iso}}_{o}\) for isotropic emission, the equation can be rewritten as

\begin{equation}
D_{L}=2.2\times10^{6}[\frac{\varepsilon \xi^{2}\kappa^{3}(1+z)^{2}}{S^{\text{iso}}_{o}(\text{Jy})\nu_{o}^{\text{iso}}(\text{GHz})(W_{o}^{\text{iso}})^{3}(\text{ms})}]^{1/2} (\text{Gpc})
\end{equation}

For typical values of \(S_{o}\), \(\nu_{o}\), \(W_{o}\) in the 33 FRBs (e.g., 1~Jy at 1~GHz, 1~ms width), \(\xi\lesssim 1\), and \(\varepsilon\sim10^{-10}\) [the efficiency of the NS kinetic energy converted to radio emission, which can be roughly estimated with the FRB luminosity in source frame \(L_{e}=4\pi D_{L}^{2}S^{\text{iso}}_{o}\nu^{\text{iso}}_{o}W^{\text{iso}}_{o}/\kappa\) and the NS kinetic energy assuming the relative NS velocity being of a few hundreds km/s (e.g., 500 km/s) when colliding with a WD], and assuming \(\kappa\sim5\) (see Ravi 2017 for some estimates), then Eq. (2) gives distances of \(<500\,\text{Gpc}\). Actually the FRB width \(W_{o}\) is defined as the full width of half maximum (FWHM) of pulse in the FRB catalog, which leads to \(\xi\lesssim 0.5\), and for the product of \(S_{o}\)\(\nu_{o}\)\(W_{o}^{3}\) (which shows a range of \(\sim1\) to about 34000, with the mean of 2082 and median value of 23), then Eq. (2) gives distances of \(<8\,\text{Gpc}\) and \(<50\,\text{Gpc}\) for the mean and median product respectively for the 33 FRBs. It is also possible that the FRBs are not isotropic emitters, and that an anisotropic factor as well as a Doppler-boost effect should be considered. Here as a measure of the anisotropy of the FRB emission, we define an effective covering factor \(\varphi\) as the ratio of the solid angle of emission to \(4\pi\) in the magnetosphere of the NS colliding with a WD. A mean Doppler factor \(\delta\) is used for the sum of coherent emission patches responsible for the FRB. Then we observe the flux density \(S_{o}(\nu)=\delta^{p} S^{\text{iso}}_{o}(\nu)/\varphi\) which relates the observed flux density of the moving plasma to the flux density that would be observed at the same frequency in the source comoving frame, where \(p=2-\alpha\) for the continuous emission, and \(p=3-\alpha\) for the discrete-jet case (Blandford \& K\"onigl, 1979; Ghisellini et al. 1993), \(\alpha\) is the radio spectral index of the FRBs defined as \(S\propto \nu^{\alpha}\). Due to the Doppler- beaming effect, the temporal width of FRB is contracted compared with that in the source comoving frame (marked ``isotropic''): \(W_{o}= W^{\text{iso}}_{o}/\delta\). With these relations, we can rewrite Eqs. (1) and (2) for the anisotropic and continuous FRB emission as

\begin{equation}
S_{o}=\frac{\varepsilon  \xi^{2}\kappa^{3}\delta^{-1-\alpha} M_{\text{NS}}D_{\text{NS}}^{2}(1+z)^{2}}{8\pi\varphi\nu_{o}D_{L}^{2}}W_{o}^{-3}
\end{equation}

\begin{equation}
D_{L}=2.2\times10^{6}[\frac{\varepsilon \xi^{2}\kappa^{3}\delta^{-1-\alpha}(1+z)^{2}}{\varphi S_{o}(\text{Jy})\nu_{o}(\text{GHz})W_{o}^{3}(\text{ms})}]^{1/2} (\text{Gpc})
\end{equation}

In the anisotropic emission, both the effective covering factor (\(\varphi<1\)) and Doppler-boost effect could enhance respectively the flux density and the distance in Eqs. (3) and (4) compared with that in Eqs. (1) and (2). For instance, for a Doppler factor \(\delta=5\), a beam covering factor \(\varphi=0.1\) (e.g., Tauris \& Manchester 1998), and \(\alpha=-2\) assumed [\(\alpha<-1\) is suggested in the CCR model by Ghisellini \& Locatelli (2018), actually spetral index is not well measured for the non-repeating FRBs (Katz 2018)], Eq. (4) gives a distance of \(\sim\)7 times that obtained from Eq. (2). However, the scaling factor \(\xi\) of FRB width here is related to the FRB beam size in an inner polar cap of a NS for the anisotropic emission, so it can be much smaller than that of the isotropic geometry. For the beam covering factor of \(\varphi=0.1\) or less (Ghisellini \& Locatelli 2018), \(\xi\) would lead to Eq. (4) gives a similar or less distance to that from Eq. (2), which could be reasonable for an extragalactic origin of the FRBs.

\begin{figure}
    \includegraphics[width=8.4cm]{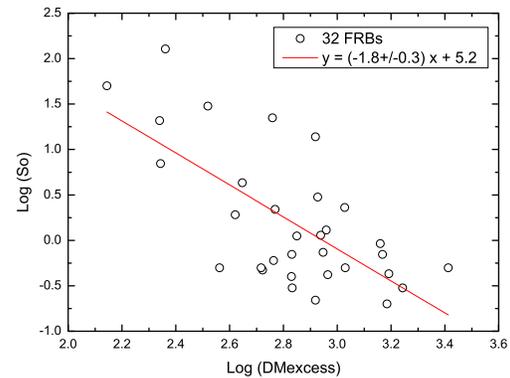}
    \caption{Log-log plot of observed peak flux density (\(S_{o}\) [Jy]) vs. \(DM_{\text{excess}}=DM-DM_{\text{Galaxy}}\) [\(\text{pc/cm}^{3}\)], for the 32 FRBs excluding the repeating one and FRB010621 which has \(DM_{\text{Galaxy}}/DM_{\text{total}}>50\%\). The best linear fit is shown.}
     \label{fig2}
  \end{figure}

The measured \(DM\)s of FRBs are significantly greater than the Galactic \(DM_{\text{Galaxy}}\) along source sightlines (except FRB010621, see below), suggesting an extragalactic origin of the FRBs. The \(DM_{\text{excess}}\) can consist of contributions from the IGM (\(DM_{\text{IGM}}\)) (Ioka 2003; Inoue 2004), the host galaxy (\(DM_{\text{host}}\)), and the source environment (\(DM_{\text{source}}\)), i.e. \(DM_{\text{excess}}=DM_{\text{IGM}}+DM_{\text{host}}+DM_{\text{source}}\). If \(DM_{IGM}\) dominates \(DM_{\text{excess}}\), we expect the observed flux density to be negatively correlated with \(DM_{\text{excess}}\) as the \(DM_{\text{IGM}}\) is proportional to the source distance. Equation (3) shows an inverse-square relation, so that \(S_{o}\propto D_{L}^{-2}\propto (DM_{\text{excess}})^{-2}\). It is roughly consistent with the statistical result with a power-law index of \(-1.8\pm0.4\) in the fit to the flux density and \(DM_{\text{excess}}\) of the 32 FRBs as shown in the log-log plot of Fig.~\ref{fig2}. Indeed this inverse square relation is simply the consequence of assuming FRBs as standard candles, whichever their intrinsic nature would be. I.e. the agreement with the observation about this relation is not an indication specifically supporting the presented model. That only indicates that, with all the assumptions about the many involved parameters (all of them spanning a small range), the face-on NS-WD merge model produces standards candles. Note that the FRB010621, which has \(DM_{\text{Galaxy}}/DM_{\text{total}}>50\%\), is excluded in the statistics of Fig.~\ref{fig2} because it is an outlier that is dominated by the Galactic DM.

\begin{figure}
    \includegraphics[width=8.4cm]{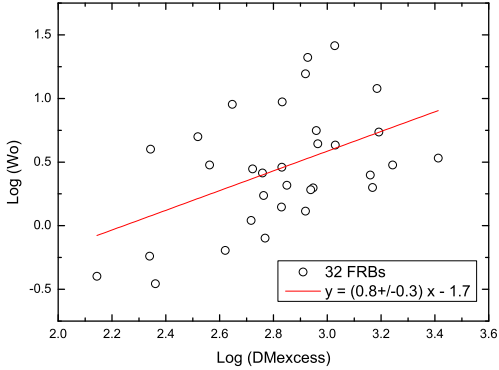}
    \caption{Log-log plot of observed pulse width (\(W_{o}\) [ms]) vs. \(DM_{\text{excess}}=DM-DM_{\text{Galaxy}}\) [\(\text{pc/cm}^{3}\)], for the 32 FRBs excluding the repeating one and the FRB010621 which has \(DM_{\text{Galaxy}}/DM_{\text{total}}>50\%\). The best linear fit is shown.}
     \label{fig3}
  \end{figure}

Furthermore, if \(DM_{IGM}\) dominates \(DM_{\text{excess}}\), from \(S_{o}\propto (DM_{\text{excess}})^{-2}\) and \(S_{o}\propto W_{o}^{-3}\), we would expect \(W_{o}\propto (DM_{\text{excess}})^{0.7}\) for a certain flux density of FRBs. This seems roughly consistent with a power-law index of \(0.8\pm0.3\) in the fit to the width and \(DM_{\text{excess}}\) of the 32 FRBs as shown in the log-log plot of Fig.~\ref{fig3}.

We note that 7 out of 34 FRBs were observed at center frequencies of 800--853 MHz, not far from the 1320--1375 MHz at which the majority FRBs are observed. This difference in frequencies should not significantly affect our statistical analysis. For instance, when we remove the seven FRBs from the sample, the slopes of the fits become \(-1.0\pm0.3\), \(-1.8\pm0.4\), and \(0.8\pm0.3\), which are similar to the slopes in Fig.~\ref{fig1}--Fig.~\ref{fig3}, respectively.

The error bars of the data in the Fig.~\ref{fig1}--Fig.~\ref{fig3} are not shown because they are not well estimated. About 27 out of 34 FRBs have measured DM errors (see, http//www.frbcat.org), they are all less than 2\% of DM values. The \(DM_{\text{excess}}\) (\(DM-DM_{Galaxy}\)), however, may have additional systematic errors for that the \(DM_{Galaxy}\) can be different from the models by Cordes \& Lazio (2002) and by Yao et al. (2017). The error of observed peak flux density of the FRBs should be larger, because this flux density is derived from observed values which could be different from that would be measured if the burst occurred at beam centre (Petroff et al. 2016). The observed pulse widths of the FRBs are obtained either with a pulse fitting algorithm or by the search code, and the errors are not given in the FRB catalog (Petroff et al. 2016). These data errors are not considered for the fits in Fig.~\ref{fig1}--Fig.~\ref{fig3}, and given the fact that there are no error bars, the fits are only tentative.


\begin{figure}
    \includegraphics[width=8.4cm]{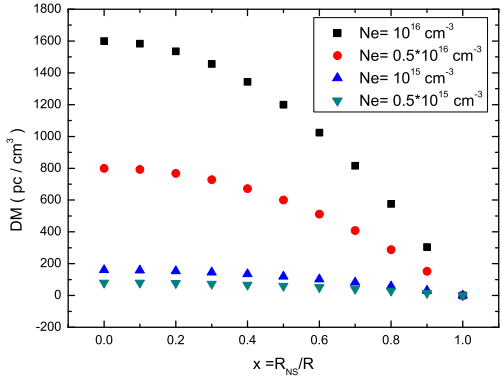}
    \caption{The DM associated to the magnetosphere of a NS, assuming the electron density decreasing outward from the NS surface as \(n_{e}=N_{e}R_{NS}^{3}R^{-3}\), and \(N_{e}\) is the electron density near the NS surface. The abscissa is the ratio of the NS radius \(R_{NS}\) and distance \(R\) from the NS centre, see the text for more explanation.}
     \label{fig4}
  \end{figure}

\section[]{Summary and discussion}

The NS-WD collision model predicts that the FRB with a shorter intrinsic width will have a higher flux density, as implied in Eqs. (1) and (3), but the scattering broadening of the width could be different from source to source due to multi-path scattering of intergalactic inhomogeneity, as discussed by Cordes \& McLaughlin (2003) and Ravi (2017). The intrinsic widths were estimated by using the models of Ravi (2017) to be \({<}1\) ms for most of the Parkes observed FRBs, which is much smaller than the observed widths. If this estimate is not model dependent and is reliable, one would expect the IGM to dominate the observed widths as well as the DMs, and so a correlation between the observed width and the \(DM_{\text{excess}}\) of the FRBs would be expected, as shown in Fig.~\ref{fig3}. It is noted by Ravi (2017) that moderate evidence exists for a relation between FRB scattering timescales and dispersion measures.

We see that the power-law index of \({\sim} -1.8\) between the flux density and \(DM_{\text{excess}}\) in Fig.~\ref{fig2} is roughly consistent with the expectations from Eqs. (1) and (3) if \(DM_{\text{excess}}\) is proportional to distance (\(DM_{\text{excess}}\propto D_{L}\)). This could be understood by rewriting Eq. (3) with \(W_{o}=\kappa W_{e}(1+z)/\delta\) as

\begin{eqnarray}
S_{o}=\frac{\varepsilon  \xi^{2}\delta^{2-\alpha} M_{\text{NS}}D_{\text{NS}}^{2}}{8\pi\varphi\nu_{o}D_{L}^{2}(1+z)}W_{e}^{-3} \\
  \propto \frac{\varepsilon  \xi^{2}\delta^{2-\alpha} M_{\text{NS}}D_{\text{NS}}^{2}}{8\pi\varphi\nu_{o}DM_{excess}^{2}(1+z)}W_{e}^{-3},
\end{eqnarray}

implying that fixing all (or most of) the model parameters in a narrow range leads to an inverse square relation alike that of Fig.~\ref{fig2}. Further analyses should be made when more FRBs are available with well-constrained intrinsic widths and distances. 

It is also useful to derive the fluence \(F\) of an FRB from Eq. (3); that is approximately, \(F=S_{o}W_{o}\propto W_{o}^{-2}\). This suggests that the shorter bursts will have a higher fluence.

It is argued that \(DM_{\text{excess}}\) could be not dominated by the IGM (Cordes et al. 2016). If this were the case, \(DM_{\text{excess}}\) would be dominated by the host galaxy and, depending on the type of host galaxy, its inclination relative to the observer, and the location of the FRB inside the galaxy (Walters et al. 2017). It would be less likely for the 32 FRBs to all be seen in edge-on host galaxies, but if all the FRBs should occur in the very central regions of host galaxies, there is a possibility of finding highly clumped gas along the line of sight. The \(DM_{\text{excess}}\) may still possibly be dominated by the IGM if \(DM_{\text{host}}\) is assumed to be \(100\ \text{pc/cm}^{3}\) (Thronton et al. 2013) or \(200\ \text{pc/cm}^{3}\) (Yang \& Zhang 2016).

Furthermore, the source environment may also have a contribution to \(DM_{\text{excess}}\). In our NS-WD collision model, a high- density plasma in the magnetosphere of the NS that was injected during collision may contribute to \(DM_{\text{excess}}\). For instance, assuming the electron density \(n_{e}=N_{e}R_{NS}^{3}R^{-3}\) (see, e.g., Sreekumar \& Schlegel 2018) in the magnetosphere of a NS, \(N_{e}\) the electron density near the NS surface, \(R_{NS}\) the NS radius, and \(R\) the radial distance from the NS centre, we can derive an integrated dispersion measure from the magnetosphere of NS along a line of sight:
\begin{eqnarray}
DM_{source}=\int^{R}_{R_{NS}}n_{e}dR=-0.5N_{e}R_{NS}^{3}(R^{-2}-R_{NS}^{-2})\\
=0.5N_{e}R_{NS}(1-x^{2}),
\end{eqnarray}
where \(x=R_{NS}/R\). Different curves of the \(DM_{\text{source}}\) from Eq. (8) are plotted for different \(N_{e}\) and \(R_{NS}=10\ \text{km}\) in Fig.~\ref{fig4}. That shows that \(DM_{\text{source}}\) can be larger than \(160\ \text{pc/cm}^{3}\) for the electron density \(N_{e}> 10^{15}\text{ cm}^{-3}\) (the median value in the Ghisellini \& Locatelli 2018 model), e.g., it can reach \(\sim800\) and \(\sim1600\ \text{pc/cm}^{3}\) for \(N_{e}=0.5\times10^{16}, 10^{16}\ \text{cm}^{-3}\) respectively. This implies that a high \(DM_{\text{source}}\) may be possible to be originated from the magnetosphere of the NS when colliding with a WD, provided that the electron density is high enough. Hence for the high \(DM_{\text{source}}\) FRBs, their redshift (that assumes \(DM_{\text{source}}=0\)) might have been over-estimated.

We should check for survival of the FRB signal across a plasma producing the DM as discussed above. In the coherent curvature model by Ghisellini \& Locatelli (2018), the plasma is mainly electron-positron pairs, which can greatly outnumber the protons. They demonstrated that the FRB emission can be optically thin around 1 GHz for the curvature self-absorption with reasonable values of parameters and a power-law distribution of electrons, e.g., for an electron density of \(N_{e}=2\times10^{15}\text{ cm}^{-3}\) which can account for a bright FRB luminosity of \(10^{43}\) erg/s, and that will lead to \(DM_{\text{source}}\) to be \(320\ \text{pc/cm}^{3}\) according to Eq. (8). The high electron density \(N_{e}=0.5\times10^{16}-10^{16}\text{ cm}^{-3}\) in Fig.~\ref{fig4} which contributes significantly to \(DM_{\text{source}}\), would produce `superluminours' FRBs with \(5\times10^{43}-10^{44}\text{ erg/s}\) still in optically thin regime at 1 GHz in the model by Ghisellini \& Locatelli (2018). However, more observations and analysis are needed to find if the `superluminours' FRBs really exist, considering that the large \(DM_{\text{source}}\) will reduce significantly the redshift estimated from total DM.

\section*{Acknowledgments}

I am grateful to the reviewer's valuable comments that have improved the paper a lot, and Z.-G. Dai for notes on the manuscript. This research is supported from the following funds: the programme of the Light in China's Western Region (grant no. XBBS201324), the 973 Program 2015CB857100, the Key Laboratory of Radio Astronomy, Chinese Academy of Sciences, and the National Natural Science Foundation of China (No.11273050).

\end{document}